\documentstyle[aps,prd]{revtex}
\begin{document}
\def \ts {\textstyle}
\def \rd {\displaystyle{\cdot}}
\def \D {\mbox{D}}
\def \ep {\varepsilon}
\def \la {\langle}
\def \ra {\rangle}
\def \c {\mbox{curl}\,}
\def \p {\partial}
\def \cs {c_{\rm s}^2}
\def \l {\Lambda}
\def \o {\Omega}
\input epsf
\def\plot#1{\centering \leavevmode
\epsfxsize= 1.0\columnwidth \epsfbox{#1}}
\newcommand{\be}{\begin{equation}}
\newcommand{\ee}{\end{equation}}
\newcommand{\ben}{\begin{eqnarray}}
\newcommand{\een}{\end{eqnarray}}
\newcommand{\n}{\label}
\newcommand{\no}{\noindent}
\newcommand{\lsim}{\mbox{\raisebox{-.3em}{$\stackrel{<}{\sim}$}}}
\newcommand{\gsim}{\mbox{\raisebox{-.3em}{$\stackrel{>}{\sim}$}}}

\title{Synergistic warm inflation}

\author{
Luis P. Chimento\footnote{Electronic address:
chimento@df.uba.ar}, Alejandro S. Jakubi\footnote{
Electronic address: jakubi@df.uba.ar}
and Norberto Zuccal\'a \footnote{Electronic address:
nzuccala@df.uba.ar}
}
\address{Departamento de F\'{\i}sica, Universidad de
Buenos Aires, 1428~Buenos Aires, Argentina}
\author{Diego Pav\'on\footnote{Electronic address:
diego@ulises.uab.es}}
\address{
Departament de F\'{\i}sica, Universidad Aut\'onoma
de Barcelona, 08193 Bellaterra, Spain}

\date{\today}

\maketitle


\begin{abstract}

We consider an alternative warm inflationary scenario in which $n$ scalar
fields coupled to a dissipative matter fluid cooperate to produce power--law
inflation. The scalar fields are driven by an exponential potential and the
bulk dissipative pressure coefficient is linear in the expansion rate. We find
that the entropy of the fluid attains its asymptotic value in a characteristic
time proportional to the square of the number of fields. This scenario remains
nearly isothermal along the inflationary stage. The perturbations in energy
density and entropy are studied in the long--wavelength regime and seen to
grow roughly as the square of the scale factor. They are shown to be
compatible with COBE measurements of the fluctuations in temperature
of the CMB.

\end{abstract}

\section{Introduction}

The very early Universe was supposedly populated by a host of
scalar fields but soon only one became to dominate the dynamics,
the others settled in the minimum of its potential. Of particular
interest from the point of view of cosmological inflation are
scalar fields with exponential potentials since these are natural
candidates to drive power--law inflation in
Friedmann--Lema\^{\i}tre--Robertson--Walker (FLRW) universes,
i.e., $ a(t) \propto t^{\alpha}$ where $a(t)$ is the scale factor
and $\alpha = \mbox{constant} > 1$ \cite{lucchin}.

As is well known scalar fields possessing exponential potentials
appear naturally in different theories of fundamental physics as
superstrings and higher dimensional theories \cite{shigher}, in
$N =2 $ supergravity \cite{salam} as well as in theories
undergoing dimensional reduction to an effective
four--dimensional theory \cite{haliwell}. However, in many cases
they happen to be too steep and fail to do the job as $\alpha
\leq 1$. Nevertheless, it has been shown that if one considers
$n$ non--interacting scalar fields with exponential potentials
they can cooperate to achieve power--law inflation in spite of the
fact that no single field can achieve it by itself
\cite{mazumdar}. The rationale behind it is that while each field
descends toward the minimum of its potential the cosmic expansion
rate (to which all the fields cooperate) acts a friction force
upon each of them. This result has been extended to Bianchi I and
VII$_{0}$ cosmologies as well \cite{juanmari}. A further
interesting feature is that the larger $n$, the closer the
resulting spectrum of initial cosmic perturbations is to
Harrison--Zeldovich's. The proponents of this scenario termed it
``assisted inflation"; however we find more fitting to call it
``synergistic inflation", and we shall do so henceforth.

A more realistic problem arises when in addition to the scalar fields one
considers a matter fluid. The price to be paid is a rather involved set of
field equations even in the simplest case when the fluid interacts with the
scalar fields only gravitationally. In a recent paper Coley and van den Hoogen
qualitatively analyzed the autonomous system of two scalar fields of the kind
discussed above in a curved FLRW universe and showed that the system has an
equilibrium point compatible with a stable phase of power--law inflation. This
feature persists even if a perfect fluid with baryotropic equation of state is
allowed in the picture \cite{alan}. Again the authors assumed that the fields
do not interact among themselves nor with the matter fluid.

At first sight the presence of a matter fluid (such as a sea of
relativistic particles) may seen as rather irrelevant because the
fast inflationary expansion will very soon dilute away these
particles. However, as shown by Berera \cite{berera} if a
coupling between the inflaton field and the matter fluid is
assumed, things change drastically to the point that when
inflation ends the Universe is far from being so cool as
predicted by ordinary inflation models. In fact it can be hot
enough to resume the radiation dominated phase of standard big--bang
cosmology, thus the reheating phase is no longer necessary.

The main target of this paper is to study the scenario of warm inflation with
$n$ scalar fields having exponential potentials and interacting with the
matter fluid. Section II considers first the more simple case of a single
scalar field plus a dissipative matter fluid, Then the analysis is extended to
$n$ scalar fields. In section III the scalar fields are allowed to decay into
the matter fluid thereby the temperature of the latter does not fall
drastically and so the warm inflationary phase can be followed by the
conventional radiation dominated period without any intermediate reheating.
Section IV studies the perturbations in energy density and entropy brought
about by warm inflation; they do not conflict with the observed temperature
anisotropies of the cosmic microwave background radiation. Finally,
section V summarizes our main findings. Units have been chosen so that
$c = 8 \pi G = 1$.\\

\section{The synergistic mechanism}

Let us assume a FLRW universe filled with a self--interacting
scalar field plus a dissipative matter fluid. The stress--energy
tensor of this mixture is

\begin{equation}
T_{ab} = (\rho+ p + \pi) u_{a} u_{b} + (p + \pi) g_{ab}, \label{2}
\end{equation}

\noindent where $\rho = \rho_{m} + \rho_{\phi}$  and $p = p_{m} +
p_{\phi}$. Here $\rho_m$ and $p_m$ are the energy density and
pressure of the matter fluid with equation of state given by
$p_{m} = (\gamma_{m} -1) \rho_{m}$ and with its baryotropic index
in the interval $ 1 \leq \gamma_{m} \le 2$. Likewise
$\rho_{\phi}$ and $p_{\phi}$, the energy density and pressure of
the minimally coupled self--interacting field $\phi$, i.e.,

\be \label{prhophi} \rho_{\phi}  =   \frac{1}{2} \dot{\phi}^{2} +
V(\phi)\, , \qquad p_{\phi} =   \frac{1}{2} \dot{\phi}^{2} -
V(\phi) \, , \ee

\no are related by an equation of state similar to that of the matter,
namely $p_{\phi} = (\gamma_{\phi} - 1) \rho_{\phi}$, so that its
baryotropic index is given by

\begin{equation} \label{gammaphi}
\gamma_\phi=\frac{\dot\phi^2}{(\dot\phi^2/2)+V(\phi)},
\end{equation}

\noindent where for non--negative potentials $V(\phi)$ one has $0
\leq \gamma_{\phi} \leq 2$, and an overdot means derivative
with respect to cosmic time. In general $\gamma_{\phi}$ varies as
the Universe expands, and the same is true for $\gamma_{m}$ since
the massive and massless components of the matter fluid redshift
at different rates.

The Friedmann equation together with the energy conservation of
the normal matter fluid with bulk dissipative pressure and
Klein-Gordon equation can be written as

\be \label{constr} \Omega_{m} + \Omega_{\phi}+\Omega_K=1,  \qquad
(K = 1, 0, -1), \ee

\be \dot{\rho_m}  +  3
H\left(\gamma_m+\frac{\pi}{\rho_m}\right)\rho_m  = 0 ,
\label{drho} \ee

\be \dot{\rho_{\phi}}+ 3H\gamma_{\phi}\rho_{\phi}= 0 , \label{KG}
\ee

\noindent where $H \equiv \dot{a}/a$ denotes the Hubble factor.
We have further introduced the density parameters $\Omega_m
\equiv \rho_{m} /\rho_{c}$, $\Omega_{\phi} ,\equiv \rho_{\phi}
/\rho_{c}$, with $\rho_{c} \equiv 3 H^{2}$ the critical density,
and $\Omega_{K} \equiv- K/(aH)^2$.

In terms of these quantities we can introduce an overall
baryotropic index~$\gamma$

\begin{equation} \label{gammaOmega}
\gamma\Omega=\gamma_m\Omega_m+\gamma_\phi\Omega_\phi \, ,
\end{equation}

\noindent where we have made use of the definition $\Omega  \equiv \Omega_{m}
+ \Omega_{\phi}$. The flatness problem is solved by the attractor solution
$\Omega=1$ of equation (\ref{constr}). In addition, the ratio
$\Omega_{\phi}/\Omega_{m}$ becomes asymptotically a constant, meaning that the
matter content of the universe does not dilutes altogether as inflation
proceeds. In \cite{lad} it was shown that these constant solutions are stable
in the $(\Omega,\Omega_m,\Omega_\phi)$ space. The fixed point solution
$\Omega=1$, $\Omega_{m} =\Omega_{m0}$ and $\Omega_\phi=\Omega_{\phi 0}$,
respectively, of equations (\ref{constr})-(\ref{KG}) is obtained when the
partial adiabatic indices and the dissipative pressure are related by

\begin{equation}
\label{gammapi} \gamma_{m}+\frac{\pi}{\rho_{m}}=\gamma_{\phi} =
-\frac{2\dot{H}}{3H^2}\, ,
\end{equation}

\noindent accordingly the smaller $\gamma_{\phi}$, the larger the
dissipative effects.

Typically bulk viscosity arises in mixtures either of different particles
species, as in a radiative fluid, or of the same species but with different
energies, as in a Maxwell--Boltzmann gas. Physically, we can think of $\pi$ as
the internal ``friction" that sets in due to the different cooling rates in
the expanding mixture. Any dissipation in exact FLRW universes has to be
scalar in nature, and in principle it may be modelled as a bulk viscosity
effect within a nonequilibrium thermodynamic theory such as the
Israel--Stewart's \cite{d1}. In that formulation and under certain
general circumstances, the evolution equation for the bulk dissipative
pressure takes the form

\begin{equation}
\pi + \tau\dot{\pi}  =  - 3\xi H \, , \label{dpi2}
\end{equation}

\noindent
where the positive--definite quantity $\xi $ stands for the
phenomenological coefficient of bulk viscosity, $T$  the temperature of the
fluid, and $\tau$ the relaxation time associated to the dissipative pressure
-i.e., the time the system would take to reach the thermodynamic equilibrium
state if the velocity divergence were suddenly turned off \cite{mdv}. Usually
$\xi$ is given by the kinetic theory of gases or a fluctuation-dissipation
theorem or both \cite{jpa}. Expression (\ref{dpi2}) has been widely used in
the literature \cite{widely} and it meets the requirements of causality and
stability to be demanded to any physically acceptable transport equation
\cite{HL}.

Combining (\ref{gammapi}) and (\ref{dpi2}) we obtain the
equation of motion of the attractor solutions satisfying
flatness, accelerated expansion and the non--dilution condition

\begin{equation} \label{dHa}
\nu^{-1}\left(\frac{\ddot H}{H}+3\gamma_m \dot H\right)+ \dot
H+\frac{3\gamma_m}{2}H^2 -\frac{3\xi}{2\Omega_{ma}}H=0 \,.
\end{equation}

\noindent Here $\nu=\left ( \tau H\right )^{-1}$ is the number of
relaxation times in a Hubble time -- for a quasistatic expansion
$\nu$ is proportional to the number of particle interactions in a
Hubble time. Perfect fluid behavior occurs in the limit
$\nu\to\infty$, and a consistent hydrodynamical description of
the fluids requires $\nu>1$.

The problem of a homogeneous scalar field driven by a exponential
potential

\be \n{v} V=V_{0}\, {\rm e}^{-A\phi} \, , \ee

\noindent minimally coupled to gravity in a flat FLRW spacetime
with a linear viscosity coefficient

\be \n{vis} \xi=\xi_{0} H \, , \ee where $V_{0}$ and
$\xi_{0}$ are constants, \noindent has the solution

\be \n{a} a=a_0 (t/t_0)^{\alpha} \ee

\be \n{fi} \phi=\phi_0\ln (t/t_0) \, . \ee

\noindent The quantities

\be \n{al} \alpha=\frac{2}{\left(1-\Omega_{ma}\right)A^2} \, , \ee

\be \n{z0} \xi_0=\Omega_{ma}\left(\gamma_m-\gamma_{\phi
a}\right) \left[1-3\gamma_{\phi a}\,\nu^{-1}\right] \, , \ee
\noindent and \be \n{gp} \gamma_{\phi a}=\frac{2}{3\alpha} \ee

\noindent are obtained by solving the system
(\ref{constr})-(\ref{KG}), (\ref{gammapi}), (\ref{dHa}) and
(\ref{v}) -the subindex $a$ stands for asymptotic value of the
corresponding quantity. The power--law expansion (\ref{a}) will be
inflationary for $A^{2} < 2 (1 - \Omega_{ma})^{-1}$.

Rewriting (\ref{dHa}) in terms of the field baryotropic  index
$\gamma_{\phi}$, we get

\be \n{gpa}
\gamma'_{\phi}=3\gamma_{\phi}^2-(\nu+3\gamma_m)\gamma_{\phi}+
\nu\left(\gamma_m-\frac{\xi}{\Omega_{ma}H}\right) \ee

\no where a prime indicates derivative with respect to $\eta=\ln a$. When the
phenomenological coefficient of bulk viscosity is given by (\ref{vis}) and
(\ref{z0}), that is $\xi_a$, Eq. (\ref{gpa}) admits the constant solution
$\gamma_\phi=\gamma_{\phi a}$. It gives an accelerated expansion in the late
time regime when $\gamma_{\phi a}<2/3$. As $\xi>0$ and $\gamma_m\ge 1$, the
hydrodynamical parameter $\nu$ is restricted to $\nu>3\gamma_{\phi a}$. The
case of constant $\xi_{0}$ arises for instance in a radiating fluid, and the
nearly linear regime, with slowly varying $\nu$ and $\gamma_m$, was already
investigated in the quasiperfect limit, corresponding to $\nu^{-1}\to 0$
\cite{lad}.

To analyze the stability of the solution
$\gamma_\phi=\gamma_{\phi a}$ we insert (\ref{vis}) in (\ref{gpa})
to obtain

\be \n{gpaa} \gamma'_{\phi}=3\left(\gamma_{\phi}^2-\gamma_{\phi
a}^2\right) -(\nu+3\gamma_m)\left(\gamma_{\phi}-\gamma_{\phi
a}\right) \ee

\no As $\gamma_{\phi a}<2/3$, $\nu>\mbox{\rm max}(3\gamma_{\phi a},1)$ and
$\gamma_m\ge 1$, Eq. (\ref{gpaa}) shows that
$\partial\gamma_\phi'/\partial\gamma_\phi<0$ in a neighborhood of
$\gamma_{\phi a}$. Hence this constant solution is asymptotically stable,
showing that all solutions of Eq. (\ref{dHa}), that is all the accelerated
attractors of the system (\ref{constr}), (\ref{drho}), (\ref{KG}),
(\ref{dpi2}), are themselves attracted towards the constant solution
$\gamma_\phi=\gamma_{\phi a}$ provided that they satisfy $\xi\sim \xi_{a}$
at late time.

\noindent
Let us now assume that instead of having just one scalar field, we
have $n$ homogeneous non-interacting scalar fields, $\phi_{i}$
with exponential potentials $V_{i}=V_{0i} \, {\rm e}^{-A_i\phi_i}$.
In that case the Einstein--Klein--Gordon equations can be recast
as

\be \label{constrn} \Omega_{m} +
\sum_{i=1}^n\Omega_{\phi_i}+\Omega_K=1,  \qquad (K = 1, 0, -1),
\ee

\be \dot{\rho_m}  +  3
H\left(\gamma_m+\frac{\pi}{\rho_m}\right)\rho_m  = 0 ,
\label{drhon} \ee

\be \dot{\rho_{\phi i}}+ 3H\gamma_{\phi i} \, \rho_{\phi i}= 0
\qquad (i = 1, 2, ... n).
\label{KGn} \ee

\noindent As it stands the problem in its full generality is rather involved.
Therefore to bring it to a form amenable to analytical treatment we shall
consider henceforth the a simplified version characterized by
$A_{1}=A_{2}=\cdots=A_n\equiv A$ and $V_{01}=V_{02}=\cdots=V_{0n}\equiv V_0$.
As we shall see in the next section, we can expect in this particular case
that all scalar fields share the same asymptotic limit. So, in the remaining
of this section we will assume $\phi_1=\phi_2=\cdots=\phi_n\equiv \phi$, so
that $V_1=V_2=\cdots=V_n\equiv V=V_0 \, {\rm e}^{-A\phi}$ and equation
(\ref{constrn}) becomes into

\be \label{constrnn} \Omega_{m} + n\Omega_{\phi_i}+\Omega_K=1,
\qquad (K = 1, 0, -1), \ee

\noindent while (\ref{KGn}) simplifies to (\ref{KG}). Following parallel steps
to that leading to Eqs. (\ref{al})-(\ref{gp}) it can be seen that the
Einstein-Klein-Gordon system has the spatially flat  power--law attractor
solution (\ref{a}) but now with

\be \n{aln} \alpha=\frac{2 \, n}{\left(1-\Omega_{ma}\right)A^2}
\, , \ee

\no showing that the $n$ scalar fields $\phi$ cooperate to a
stronger inflation.

Let us now assume that the $n$ homogeneous scalar fields,
$\phi_i$ are driven by a general potential
 $V=V(\phi_i)$. In
that case the Einstein-Klein-Gordon equations
 are

\begin{equation}
\label{b001} 3H^{2}=
\frac{1}{2}\sum_{i=1}^n\dot\phi_i^{2}+V+\rho_m- \frac{3K}{a^2},
\end{equation}

\begin{equation}  \label{bkg1}
\ddot{\phi_i}+3H\dot{\phi_i}+V_{,\phi_i}=0.
\end{equation}

\no along with equation (\ref{drhon}) ($V_{,\phi_i}$ stand for
$\partial V/\partial {\phi_i}$). From these equations we get

\begin{equation}  \label{bhp}
\dot{H}=-\frac{1}{2}\sum_{i=1}^n \dot{\phi_i}^{2}-
\frac{1}{2}(\gamma_m\rho_m+\pi)+\frac{K}{a^2}.
\end{equation}

In order to investigate the stable scalar field configurations
it is expedient to introduce the ancillary quantity

\begin{equation} \label{om}
\omega=\frac{\sum_{i=1}^n \dot{\phi_i}^{2}}{n\dot\phi_{\alpha}^2},
\end{equation}

\no which reduces to $\omega=1$ for the completely symmetric configuration
$\phi_1=\phi_2=\cdots=\phi_n$.  Using (\ref{b001})--(\ref{om}) we
find the differential equation for $\omega$:

\begin{equation} \label{dom}
\dot\omega=2\frac{nV_{,\phi_\alpha}\dot\phi_{\alpha}\omega-\dot
V}{n\dot\phi^2_{\alpha}}.
\end{equation}

\no (In this section no summation convention applies to repeated
Greek indices). If we further assume that the potential satisfies
the condition

\begin{equation} \label{poa} \dot
V=nV_{,\phi_\alpha}\dot\phi_{\alpha},
\end{equation}

\no then equation (\ref{dom}) becomes

\begin{equation} \label{doma}
\dot\omega=2\frac{V_{,\phi_\alpha}}{\dot\phi_{\alpha}}(\omega-1)
.
\end{equation}

\no This has a fixed point solution, namely $\omega=1$. Further;
with the aid of (\ref{bhp})-(\ref{om}) the
general solution of (\ref{doma}) can be found in terms of the
scale factor, the matter fluid and the bulk viscous pressure

\begin{equation} \label{omg}
\omega=\left(1+\frac{nc}{2a^6\left[\dot H+
\frac{1}{2}(\gamma_m\rho_m+\pi)-\frac{K}{a^2}\right]}\right)^{-1},
\end{equation}

\no where $c$ is an arbitrary integration constant. Evaluating
(\ref{omg}) in the asymptotic regime of the attractor
solution $\gamma_\phi=\gamma_{\phi a}$ and the potentials
$V_{i}=V_{0i}{\rm e}^{-A_i\phi_i}$, which asymptotically
satisfies the condition (\ref{poa}), it can be easily shown that
the particular solution $\omega=1$ is an attractor for evolutions
that behave asymptotically as $a\propto t^\alpha$ with
$\alpha>1/3$. On the other hand this result strongly suggests
that the special case in which all scalar fields are equal may be
the late-time attractor of more general scenarios.

\section{Warm inflation}

Warm inflation arises when a strong enough coupling between
the scalar field and matter fluid (which we shall assume perfect) exists.
The former decays into the latter (which acts as a thermal bath) while the inflaton
slowly rolls down the potential \cite{berera}. The decay is
phenomenologically implemented by inserting a (usually constant)
friction term $\Gamma$ in the equation of evolution for $\phi$

\be\label{dphiGamma} \ddot\phi + 3(H + \Gamma)\dot\phi + V'(\phi)= 0. \ee

\noindent We adopt this picture except that (i) we consider no slow--roll
(although it can be straightforwardly incorporated), and (ii) rather than a
single field we have $n$ scalar fields all of them with identical exponential
potential. Here $\phi$ stands for any of these fields. Accordingly
(\ref{dphiGamma}) is not just a single equation but $n$ identical equations;
besides for mathematical simplicity we assume the same friction term for each
field.\\

\noindent
Obviously the coupling between the $n$ scalar fields and the matter fluid
introduces a source term in the energy balance equation for the latter

\be\label{drhoGamma} \dot\rho_m + 3\gamma_m H \rho_m = 3\Gamma
\dot\phi^2. \ee

\noindent So there is a continuous transfer of energy from the scalar fields
to the matter adjusted in such a way that the former experience a damped
evolution and give rise to a nearly isothermal expansion. Accordingly, like in
standard warm inflation, no reheating mechanism is needed at the end of
inflation. Moreover, thermal rather than quantum fluctuations produce the
primordial spectrum of density perturbations
\cite{Bererafang,Taylor,Leefang,Bellini,BGR}.

We can identify the phenomenological coupling with an effective
dissipative pressure $\pi^*$ along the evolution on the attractor.
Then comparing (\ref{drho}) with (\ref{drhoGamma}) we get

\begin{equation} \label{Gammapi}
\pi^*=-\frac{\Gamma\dot\phi^2}{H} \,.
\end{equation}

\noindent
In the case at hand the attractor condition (\ref{gammapi}) becomes
dynamic because the starred magnitudes include the interaction between the
scalar field and radiation

\begin{equation} \label{newattractor}
\gamma_{m} + \frac{\pi^*}{\rho_{m}} = \gamma^*_{\phi}
= - \frac{2 \dot{H}}{3H^{2}} \, ,
\end{equation}
where
$\gamma^*_{\phi}\equiv\gamma_{\phi}-(\pi^*/\rho_{\phi})  $.
Using this together with (\ref{prhophi}) and (\ref{gammaphi}) it follows
that

\begin{equation} \label{GammaH}
\Gamma=\left(\frac{\gamma_m}{\gamma^*_{\phi}}-1\right)
\frac{\rho_m}{\rho_\phi} H\equiv R H
\end{equation}

\noindent (bear in mind that $\gamma_{m} > \gamma^*_{\phi}$).
\noindent This implies that the scalar field evolves with an
``effective" expansion rate $\tilde H =(1+R)H$. Once converged to
the attractor solution, $R$ becomes a constant and the
effective power--law exponent results larger by a factor $(1+R)$:

\begin{equation} \label{alphanR}
\alpha=\frac{2n\left(1+R\right)}{\left(1-\Omega_{ma}\right)A^2} .
\end{equation}

\noindent
This means that this interaction between the scalar fields and
radiation further assist to inflation.

Let us investigate the entropy production $\dot S$ along this era. Using
(\ref{newattractor}) and (\ref{drho}) we get that matter redshifts as
$\rho_m=\rho_{me}(a/a_e)^{-3 \gamma^*_\phi}$ (subindex $e$ means evaluation
at inflation exit). As it is customary in warm
inflationary scenarios we assume that this matter behaves as radiation
$\rho_m=(\pi^2/30)gT^4$, where $g$ is the effective number of
relativistic degrees of freedom, so we have

\begin{equation} \label{Ta}
T=T_e\left(a/a_e\right)^{-3\gamma^*_\phi/4}.
\end{equation}

\noindent
Then, using (\ref{gammapi}), (\ref{vis}), (\ref{Ta}), (\ref{a}) and
$\dot S=\pi^{*2}/\xi T$, we get

\begin{equation} \label{St}
S(t)=S_e\left[1-\left(\frac{\tau_1}{t}\right)^{3/2}\right].
\end{equation}

\noindent
where

\begin{equation} \label{tau1}
\tau_1=\left[
\frac{16\left(\gamma^*_\phi-4/3\right)^2\Omega_{ma}^2}
{9{\gamma^*_\phi}^3\xi_{0}T_e S_e t_e^{1/2}}\right]^{2/3}.
\end{equation}

\noindent is the characteristic time for the entropy density to attain a
constant value $S_e$ and is directly related to the time for the start of warm
inflation.

We see from (\ref{Ta}) that the requirement of a nearly
isothermal expansion along the inflationary era imposes a
constraint on the value of $\gamma^*_\phi$. Namely, if $T_{i}$ is
the temperature of the radiation fluid at the beginning of
inflation, $T_e$ at the end, and $N$ is the number of e-folds,
we have

\begin{equation} \label{Ti0}
\frac{T_i}{T_e}= e^{-3\gamma^*_\phi N/4}.
\end{equation}

\noindent Assuming $T_{i}/T_e = {\cal O}(1)$ we find that $\gamma^*_\phi\simeq
10^{-2}$. From (\ref{gp}) and (\ref{aln}) we see that this can be easily
achieved, with $n$ in the range of $10-100$, without fine--tuning the
potential slope.

Inflationary scenarios need to achieve a graceful exit from inflation. In our
case this is not a problem since the ratio $\rho_\phi/\rho_m$ is not exactly
constant,  as can been seen from (\ref{dphiGamma}) and (\ref{drhoGamma}).
Therefore, the continuous transfer of energy from the $n$ scalar fields to the
matter fluid slowly increases the energy of the latter and decreases that of
the former. Thus the acceleration equation $\ddot{a}/{a} =
-\textstyle{1\over{6}}(\rho + 3p)$ implies that the universe ceases to
accelerate when both energy densities equalize. This criterion for ending
inflation coincides with Taylor and Berera's \cite{Taylor}. \\

\section{Evolution of perturbations}

This section considers the evolution of energy density, entropy and curvature
fluctuations in the perturbative long--wavelength regime during the attractor
era. Scalar perturbations are covariantly and gauge--invariantly characterized
by the spatial gradients of scalars. Energy density inhomogeneities are
described by the comoving fractional density gradient \cite{eb}

\begin{equation}
\delta _{i}={\frac{a\D_{i}\rho }{\rho }} \,,  \label{e'}
\end{equation}

\noindent where $\D_{i}$ stands for the covariant spatial derivative $\D
_{j}A_{i\cdots }=h_{j}{}^{k}h_{i}{}^{l}\cdots \nabla _{k}A_{l\cdots }$. The
scalar part $\delta \equiv a\D^{i}\delta _{i}=(a\D)^{2}\rho /\rho $ encodes
the total scalar contribution to energy density inhomogeneities. It relates to
the usual gauge-invariant density perturbation scalar $\ep_{ \rm m}$ through
$\delta=\nabla^2 \ep_{ \rm m}$, where $\nabla^2$ is the Laplacian for the
metric of the 3--surfaces of constant curvature \cite{b,bde}. Also the
comoving expansion gradient, the normalized pressure gradient, and normalized
entropy gradient are defined by \cite{eb,mt}

\begin{equation}
\theta _{i}=a\D_{i}\theta \,,\quad p_{i}={\frac{a\D_{i}p}{\rho }}
\,,\quad e_{i}={\frac{a n T\D_{i}s}{\rho }}\, ,  \label{e''}
\end{equation}

\noindent  $n$ being the particle number density, $T$ the
temperature, and $s$ the specific entropy per particle. The
evolution equation for scalar density perturbations reads
\cite{mt}

\[
\ddot{\delta}+H\left( 8-6\gamma +3\cs\right)
\dot{\delta}-{\ts{3\over2}} H^{2}\left\{ -10+14\gamma-3 \gamma
^{2}-6\cs+\right.
\]
\begin{equation}
\left.\left[( 1-3\left( \gamma -1\right) ^{2}+2\cs\right]
k\right\} \delta -\cs\D ^{2}\delta ={\sf S}[e]+{\sf S}[\pi^*]+{\sf
S}[q]+{\sf S} [\sigma ]\,,  \label{l}
\end{equation}

\noindent where

\begin{equation}  \label{cs}
\cs=\left( {\frac{\p p}{\p\rho }}\right) _{\!s} \,,\quad
r={\frac{1}{nT}}\left( \frac{\p p}{\p s}\right) _{\!\rho },
\end{equation}

\noindent are, respectively, the adiabatic speed of sound and a
non-baryotropic index. The sources in the right--hand side of Eq.
(\ref{l}) arising, respectively, from entropy perturbations, bulk
viscous stress, energy flux, and shear viscous stress are given
in \cite{mt}. Since in our case there are no shear viscous stress
($\sigma _{ij}=0$) and ${\sf S}[q]$ vanishes by choosing the
energy frame ($q_{i}=0$), we reproduce here only the expressions
for $ {\sf S}[e]$ and ${\sf S}[\pi^*]$:

\begin{eqnarray}
{\sf S}[e] &=&r\left( 3K H^{2}+\D^{2}\right) e\,,  \label{l1} \\
{\sf S}[\pi^*] &=&-\left( 3K H^{2}+\D^{2}\right) {\cal B}\,,
\label{l2}
\end{eqnarray}

\noindent where the scalar entropy perturbation

\begin{equation}
e=a\D^{i}e_{i}={\frac{a^{2}nT}{\rho }}\D^{2}s
\end{equation}

\noindent and the dimensionless perturbation scalar

\begin{equation}
{\cal B}={\frac{a^{2}\D^{2}\pi^{*} }{\rho}}\,,~  \label{l5}
\end{equation}

\noindent related to the inhomogeneous part of the bulk viscous stress, were
defined. Also, the entropy perturbation equation in the energy frame is

\begin{equation}
\dot{e}+3H\left( \cs-\gamma +1+r\right) e=-3H {\cal B}\ .
\label{entr}
\end{equation}

The coupled system that governs scalar dissipative perturbations in the
general case is given by the energy density perturbation equation (\ref{l}),
the entropy perturbation equation (\ref{entr}), the equation for the scalar
bulk viscosity (\ref{dpi2}), and the equation for temperature perturbations.

When only bulk viscous stress dissipation is present, the coupled system can
be reduced to a pair of coupled equations in $\delta $ (third order in time)
and $e$ (second order in time). For a flat background, the equations are
\cite{mt}:

\[
\tau  \stackrel{\ldots }{\delta }+\left[ 1+3\left( 2-\gamma
+\cs\right) \tau  H\right] \ddot{\delta}+H\left\{ 8-6\gamma
+3\cs+3\tau  \left( \cs \right) ^{\rd}\right.
\]
\[
\left. -{\ts{1\over2}}\left[ -14+75\gamma -48\gamma
^{2}+(21\gamma -30)\cs \right] \tau  H\right\} \dot{\delta}
\]
\[
-{\ts{3\over2}}H^{2}\left\{ 6-10\gamma +5\gamma ^{2}-6\cs-4\tau
\left( \cs \right) ^{\rd }\right.
\]
\[
\left. -2\left[ -6+18\gamma -15\gamma ^{2}+5\gamma ^{3}+\left(
6-28\gamma +10\gamma ^{2}\right)  \cs\right] \tau H\right\} \delta
\]
\[
={\frac{a^{2}\xi }{\rho  \gamma }}\D^{2}\left(
\D^{2}\dot{\delta}\right) + {\frac{3a^{2}(\gamma -1)H}{\rho
\gamma }}\D^{2}\left( \D^{2}\delta \right) +\tau
\cs\D^{2}\dot{\delta}+\tau r\D^{2}\dot{e}
\]
\[
+\left[ \left( 1-3\gamma  \tau  H\right) \cs+\tau \left( \cs
\right) ^{\rd}+3\left( {\frac{\p\xi }{\p\rho }}\right)
_{\!s}\right] \D ^{2}\delta
\]
\begin{equation}
+\left[ \left( 1-3\gamma \tau  H\right) r+\tau  \dot{r}+3\left(
{\frac{ \p\xi }{\p s}}\right) _{\!\rho }\right] \D^{2}e\,,
\label{d2} \end{equation}

\noindent and

\[
\tau  \ddot{e}+\left[ 1-{\ts{3\over2}}\left( -2+3\gamma
-2\cs-2r\right) \tau  H\right] \dot{e}
\]
\[
-3H\left[ \gamma -1-\cs-r+3\gamma  \left( \gamma -\cs\right)
\tau  H+\tau
 \left( \cs
+r\right) ^{\rd}\right.
\]

\begin{equation}
\left. -{\frac{\rho }{\gamma }}\left( {\frac{\p\xi }{\p
s}}\right) _{\!\rho }\right] e=-{\frac{\xi }{\gamma }\
}\dot{\delta}+{\frac{3H}{ \gamma }}\left[
\left(\gamma-1\right)\xi +\rho \left( {\frac{\p\xi }{\p\rho
}}\right) _{\!s}\right] \delta \,.  \label{d3}
\end{equation}

We shall consider here the evolution of the energy density and entropy
perturbations in the attractor stage with the conditions
$r=0$ and $\partial \nu /\partial s=0$. Together with Eq.
(\ref{vis}) they imply

\begin{eqnarray}
\left( \frac{\partial \xi }{\partial s}\right) _{\rho } &=&0\ , \\
\left( \frac{\partial \xi }{\partial \rho }\right) _{s}
&=&\left( \frac{
\partial \xi }{\partial \rho _{m}}\right) _{s}=\frac{ \xi_{0} }
{6\Omega _{ma}H}.
\end{eqnarray}

\noindent
In this case Eq. (\ref{d2}) decouples to give
\begin{eqnarray}
\stackrel{\ldots }{\delta
}+\frac{c_{1}}{t}\ddot{\delta}+\frac{c_{2}}{t^{2}}
\dot{\delta}+\frac{c_{3}}{t^{3}}\delta &=&c_{4}\ t^{2\alpha }\
\D^{4}\dot{ \delta}+c_{5}\ t^{2\alpha }\ \D^{4}\delta +\cs
\D^{2}\dot{\delta}
\nonumber \\
&&+\left( \frac{c_{6}}{t}+c_{7}\right) \D^{2}\delta \ ,
\label{delta}
\end{eqnarray}

\noindent and Eq. (\ref{d3}) becomes
\begin{equation}
\ddot{e}+\frac{c_{8}}{t}\ \dot{e}+\frac{c_{9}}{t^{2}}\
e=\frac{c_{10}}{t^{2}} \ \dot{\delta}+\frac{c_{11}}{t^{2}}\
\delta  \label{entrop}
\end{equation}

\noindent where the constant coefficients $c_{1}\ldots c_{11}$ depend upon the
parameters of the model: $\nu $, $\kappa _{1}$, $\kappa _{2}$, $\Omega
_{m},\alpha $, $\gamma $, $\cs$, and the value of the scale factor $a_e$ at
the exit from inflation. For our purposes, only the explicit expression for
$c_{1}$, $c_8$, and $c_9$ are relevant, being

$$
c_{1}=\alpha \nu +3\alpha \ \left( 2-\gamma -\cs \right) \,,
$$
$$
c_8=\alpha\left[\nu-\frac{3}{2}\left(3\gamma-2-2\cs\right)\right]
\,,
$$
\begin{equation}\label{c1}
c_9=-3\alpha^2\left[\nu\left(\gamma-1-\cs\right)+
3\gamma\left(\gamma-\cs\right)\right] \,.
\end{equation}

\noindent We deal with the system (\ref{delta}),(\ref{entrop}) by
performing separation of variables  in the form $\delta =\delta
_{x}\ \delta _{t}$ and $ e=e_{x}\ e_{t} $, where $\delta _{x}$
and $e_{x}$ depend upon the spatial variables while $\delta _{t}$
and $e_{t}$ are functions of the coordinate time $t.$ Then, Eq.
(\ref{delta}) can be recasted as

\begin{eqnarray}
\frac{\stackrel{\ldots }{\delta }_{t}}{\delta
_{t}}+\frac{c_{1}}{t}\frac{ \ddot{\delta}_{t}}{\delta
_{t}}+\frac{c_{2}}{t^{2}}\frac{\dot{\delta}_{t}}{ \delta
_{t}}+\frac{c_{3}}{t^{3}} &=&t^{2\alpha }\left( c_{4}\frac{\dot{
\delta }_{t}}{\delta _{t}}+c_{5}\right) \frac{\D^{4}\delta
_{x}}{\delta _{x}}
\nonumber\\
&&+\left( \cs \frac{\dot{\delta}_{t}}{\delta _{t}}+\frac{c_{6}}{t}
+c_{7}\right) \frac{\D^{2}\delta _{x}}{\delta _{x}}\ ,
\end{eqnarray}

\noindent which can only hold if

\begin{equation}
\left( \D^{2}-\mu \right)  \delta _{x}=0\ ,  \label{proca}
\end{equation}

\noindent  $\mu $ being an arbitrary constant. In Fourier transformed space it
becomes an identity that holds for an arbitrary amplitude $\delta_k$ with
$\mu=-k^2/a^2$, where $k/a$ is the physical wavenumber. Also, Eq.
(\ref{entrop}) leads to

\begin{equation} \label{deltae}
\frac{t^{2}\left( \ddot{e}_{t}/e_{t}\right) +c_{8}\ t\ \left(
\dot{e} _{t}/e_{t}\right) +c_{9}}{c_{10}\left(
\dot{\delta}_{t}/e_{t}\right) +c_{11}\left( \delta
_{t}/e_{t}\right) }=\frac{\delta _{x}}{e_{x}}\ ,
\end{equation}

\noindent which requires $e_{x}=A\ \delta _{x}\ $, with $A$ a constant. Then
the evolution equation for mode $\mu$ becomes

\[
\stackrel{\ldots }{\delta
}_{t}+\frac{c_{1}}{t}\ddot{\delta}_{t}+\left(
\frac{c_{2}}{t^{2}}-\mu ^{2}\ t^{2\alpha }\ c_{4}-\mu \cs \right)
\dot{ \delta }_{t}
\]
\begin{equation}  \label{deltat}
+\left[ \frac{c_{3}}{t^{3}}-\mu ^{2}\ t^{2\alpha }\ c_{5}-\mu
\left( \frac{ c_{6}}{t}+c_{7}\right) \right] \delta _{t} =0\ .
\end{equation}

As we are interested in the asymptotic behavior of the perturbations along the
attractor regime, it suffices to consider the dominant terms in (\ref{deltat})
for large time. As $\alpha>0 $ we have

\begin{equation}
\stackrel{\ldots }{\delta
}_{t}+\frac{c_{1}}{t}\ddot{\delta}_{t}-\mu ^{2}\ t^{2\alpha }\
c_{4}\ \dot{\delta}_{t}-\mu ^{2}\ t^{2\alpha }\ c_{5}\ \delta
_{t}\cong 0\ .  \label{asymptotic}
\end{equation}

\noindent
To study the asymptotic evolution of the long--wavelength modes much larger
than the Hubble scale ($k/aH\ll 1$), we expand $\delta _{t}$ in powers of $\mu
^{2}$ as $\delta _{t}\cong \delta _{t}^{(0)}+\mu ^{2}\ \delta _{t}^{(1)}$.
Then, replacing this expression in Eq. (\ref{asymptotic}) and retaining terms
up to first order in $\mu ^{2}$, we obtain

\begin{equation}\label{deltat1}
\left[ \stackrel{\ldots }{\delta
}_{t}^{(1)}+\frac{c_{1}}{t}\ddot{\delta} _{t}^{(1)}-t^{2\alpha }\
c_{4}\ \dot{\delta}_{t}^{(0)}-t^{2\alpha }\ c_{5}\ \delta
_{t}^{(0)}\right] \ \mu ^{2}+\stackrel{\ldots }{\delta
}_{t}^{(0)}+ \frac{c_{1}}{t}\ddot{\delta}_{t}^{(0)}=0\ .
\end{equation}

\noindent Its zeroth--order solution is

\begin{equation}
{\delta}_{t}^{(0)}(t)=\frac{A_{1}}{(1-c_{1})(2-c_{1})}\
t^{2-c_{1}}+A_{2}\ t+A_{3}\ ,  \label{delta0}
\end{equation}

\noindent for $c_{1}\notin \{1,2\}$ and being $A_{i}$ , $i=1,2,3$
arbitrary integration constants. Then, $\delta _{t}^{(1)}$ satisfies the
inhomogeneous equation

\begin{equation} \label{deltat2}
\stackrel{\ldots }{\delta }_{t}^{(1)}+\frac{c_{1}}{t}\ddot{\delta}
_{t}^{(1)}-t^{2\alpha }\ c_{4}\ \dot{\delta}_{t}^{(0)}-t^{2\alpha
}\ c_{5}\ \delta _{t}^{(0)}=0\ ,
\end{equation}

\noindent whose general solution has the form

\begin{eqnarray}
\delta _{t}^{(1)} &=&\frac{B_{1}}{(1-c_{1})(2-c_{1})}\
t^{2-c_{1}}+B_{2}\
t+B_{3}+a_{1}t^{2\alpha +4-c_{1}} \nonumber\\
&&+a_{2}\ t^{2\alpha +3}+a_{3}\ t^{2\alpha +5-c_{1}}+a_{4}\
t^{2\alpha +4}\ ,
\end{eqnarray}

\noindent where the coefficients $B_{i}$, $i=1,2,3$ are integration constants
and $a_{i}$, $i=1,\ldots ,4$ depend on $\alpha $, $c_{1}$, and the $B_{i}$.
Bearing in mind that $\alpha \gg 1$, we find that the dominant mode of the
energy density perturbations in the super--Hubble regime grows asymptotically
like $t^{2\alpha+4}\propto a^{2+(4/\alpha)}\simeq a^{2}$, independently of the
wavenumber, until inflation exit.

On the other hand, Eq. (\ref{d3}) becomes, in the leading regime,

\begin{equation} \label{de}
\ddot e_t+\frac{c_8}{t}\dot e_t+\frac{c_9}{t^2}e_t=
c_{12}t^{2\alpha+1}\left[1+\frac{3}{\nu}\left(\gamma-1+
\frac{1}{2\Omega_{ma}}\right)t\right] \,,
\end{equation}

\noindent whose solution is

\begin{equation} \label{et}
e_t(t)=B_4 t^{\lambda_1}+B_5 t^{\lambda_2}+c_{13}
t^{2\alpha+3}\left[1+\frac{3}{\nu}\left(\gamma-1+
\frac{1}{2\Omega_{ma}}\right)t\right] \,,
\end{equation}

\noindent where $\lambda_{1,2}$ are the roots of the equation
$\lambda^2+(c_8-1)\lambda+c_9=0$, $B_4, B_5$ are arbitrary integration
constants, and $c_{12},c_{13}$ are functions of the parameters and the
previously defined integration constants. It follows that the dominant mode of
the entropy perturbations also grows as $a^{2+(4/\alpha)}\simeq a^2$ for
$\nu={\cal O}(1) $.

To deal with the evolution of the curvature perturbations generated by these
energy density and entropy fluctuations we will turn to the standard
metric--based gauge--invariant approach. In a particular choice for slicing of
space--time named the longitudinal gauge, the metric describing the
inhomogeneous perturbations of the spatially flat FLRW background takes the
simple form

\begin{equation}
d s^2 = (1+2 \Phi)\,d t^2 - a^2(t) (1-2 \Psi) \delta_{i j}\,d x^i\,d x^j,
\label{eq15}
\end{equation}

\noindent in terms of the gauge invariant Bardeen potentials $\Phi$ and $\Psi$
\cite{b,MFB}. Besides, when the  shear stress vanishes, it follows from the
equations of motion for the gauge invariant variables that $\Phi = \Psi$ . So,
just a single scalar degree of freedom, say $\Phi$, is required to describe
linear perturbations of the metric. We get two second order equations,
namely

\begin{equation} \label{lPhi}
\nabla^2 \Phi - 3 a H \Phi^\prime - 3 a^2 H^2 \Phi
= \frac{1}{2}a^2 \delta \rho\,,
\end{equation}

$$
\Phi^{\prime\prime} + 3{a H} \left (1 + c_s^2\right ) \Phi^\prime -
c_s^2 \nabla^2 \Phi + \left [ 2a \left(a H\right)^\prime + (1 + 3c_s^2 ) a^2
H^2 \right ]\Phi
$$

\begin{equation} \label{dPhi}
= \frac{1}{2}a^2 \left(\delta p+\delta\pi^*-c_s^2\delta\rho\right),
\end{equation}

\noindent where the prime denotes derivative with respect to conformal time,
and the source terms contain the energy density perturbation $\delta\rho$, the
equilibrium pressure perturbation $\delta p$ and the dissipative pressure
perturbation $\delta\pi^*$.

When the source term of Eq. (\ref{dPhi}) vanishes, and the
scales are larger than the Hubble radius such that the spatial
gradients can be neglected, Eq. (\ref{dPhi}) can be
recast in terms of the curvature perturbation on uniform--density
hypersurfaces \cite{BST,Bardeen88,MS98}

\begin{equation} \label{zeta}
\zeta \equiv \Phi+H\frac{\delta\rho}{\dot\rho}=
\Phi+\frac{2}{3}\frac{H^{-1} {\dot \Phi} + \Phi}{\gamma+\pi^*/\rho}
\end{equation}

\noindent as a conservation law ${\dot \zeta} = 0$ \cite{WMLL,gordon}. We
consider that the equilibrium pressure perturbation is isentropic during warm
inflationary and radiation dominated eras, and that the dissipative pressure
perturbation switches off during the transition between both eras with a
relaxation time that is a fraction of a Hubble time. Hence, one should expect
that $\dot\zeta$ soon vanishes along this transition so that the value
$\zeta_e$ at inflation exit on super--Hubble scales may be equated to that at
reentry of long wavelength modes to Hubble radius during the radiation-- or
matter--dominated eras. To find this value we need the evolution of $\Phi$
along the attractor era for long--wavelength modes. Again, neglecting the
spatial derivative term and using the relationship
$\epsilon_{mk}=-\delta_t\delta_k/k^2\simeq -a^2\delta_k/k^2$, the Fourier
transform of Eq. (\ref{lPhi}) becomes

\begin{equation} \label{dPhik}
a\frac{d\Phi_k}{da}+\Phi_k=\frac{\delta_k}{2k^2}a^2.
\end{equation}

\noindent
Its solution

\begin{equation} \label{Phik}
\Phi_k=\frac{C_1}{a}+\frac{\delta_k}{6k^2}a^2
\end{equation}

\noindent shows that $\Phi$ grows asymptotically as $a^2$ during the attractor
regime. Inserting this result back in (\ref{zeta}) we find that $\zeta$ also
grows as $a^2$ during this regime. This nonconservation of the super--Hubble
curvature perturbations is a consequence of the growth of the entropy
perturbations, which, in its turn, is due to dissipation effects as
Eq.(\ref{entr}) shows. Then, taking into account that $\gamma_\phi^*\ll 1$, we
find that $\zeta\simeq 3\alpha \Phi\simeq -\alpha\epsilon_m/2$ at inflation
exit. Hence the power spectra of $\zeta$ and $\Phi$ at that moment are
proportional to the power spectrum of the primordial energy density
perturbations.

For a mode that crosses outside the Hubble radius at scale factor $a_A$ during
inflation and reenters to the Hubble radius at scale factor $a_B$ during the
radiation dominated era, we find the number of e--foldings before the end of
inflation $N_A=\alpha/(\alpha-1)\ln(a_B/a_e)$ by using the continuity of the
energy density at the turnover between the warm inflation and the radiation
dominated eras. Then, for this mode of perturbations, the regime of growth as
$a^2$ starts at scale factor $a_1\gg a_A$, when both $aH\gg k$ and the
evolution is close to the attractor, and it continues until inflation exit.
Then we obtain

\begin{equation} \label{growthr}
\frac{a_e}{a_1}=\frac{1}{\sigma}\left(\frac{a_B}{\beta a_N}\right)
^{\alpha/(\alpha-1)}
\simeq \frac{a_B}{\beta\sigma a_N}
\end{equation}

\noindent where $\sigma=a_1/a_A$, $a_N$ is the scale factor at the start of
nucleosynthesis era, corresponding to a temperature $T_N\simeq 1$MeV,
$\beta=a_e/a_N$ and we are taking $\alpha\gg 1$. As in this warm inflationary
scenario there is no need to accommodate a reheating stage, inflation may end
shortly before nucleosynthesis and we may take safely $\beta\simeq 10^{-2}$.
Then we obtain an upper bound on the growth of perturbation modes crossing
inside the Hubble radius during the radiation dominated era by considering the
matter--radiation equality scale $k_{eq}^{-1}\simeq 100$Mpc, corresponding to
a cluster of galaxies. As $a_{eq}/a_N\simeq 10^6$, taking $\sigma\gsim 10^3$
we find that $\epsilon_{me}/\epsilon_{m1}=\zeta_e/\zeta_1\lsim 10^{10}$.
Similarly, for a mode crossing inside the Hubble radius during the matter
dominated era, we obtain

\begin{equation} \label{growthm}
\frac{a_e}{a_1}\simeq \frac{3}{4\beta\sigma}
\left(\frac{a_{eq}}{a_N}\right)\left(\frac{a_B}{a_{eq}}\right)^{1/2}
\end{equation}

\noindent Thus, scales $k_{hor}^{-1}\simeq 10^4$Mpc, corresponding to the
observable universe, give the upper bound on the growth of perturbation modes
$\zeta_e/\zeta_1\lsim 10^{13}$.

After the inflationary stage, we recover the standard picture of conserved isentropic
curvature perturbations and well known calculations show that the density
contrast at Hubble scale entry $(\delta\rho/\rho)_{k=aH}$ is proportional to
$\zeta$, or equivalently to the comoving curvature perturbation $\cal R$, with
a proportionality factor of $2/5$ during presureless matter dominated era, or
$4/9$ during the radiation dominated era \cite{LR}. So, using COBE
normalization for the power spectrum of curvature $(2/5){\cal P}_{\cal
R}^{1/2}=1.91\times 10^{-5}$ at the scale $k^{-1}\simeq 10^3$Mpc, we find that
at inflation exit $\epsilon_{mk}=9.55\times 10^{-5}/\alpha\simeq 10^{-6}$ for
$\alpha\simeq 100$. Besides, due to the proportionality of power spectra, the
observed bound on the spectral slope of curvature perturbations $n=1.0\pm 0.2$
implies the same bound on the spectral slope of the energy density
perturbations at inflation exit. Recalling that the amplitude $\delta_k$ is an
arbitrary function, these observational bounds impose no fine tuning
constraints whatsoever on the parameters of the scalar field potential.

\section{Discussion}

We have proposed a new inflationary scenario whose main ingredients are $n$
scalar fields and a dissipative matter fluid. The former decay into the
latter at a high rate $\Gamma$. While no single scalar field can
achieve inflation by its own they cooperate synergistically to produce it. We
have derived the attractor condition Eq. (\ref{gammapi}) and shown that the
presences of dissipation does not spoil the linear relationship between the
power--law exponent $\alpha$ and the number of fields, preserving the
stability of the symmetric configuration of $n$ identical fields.

We have described the interaction between the scalar fields and the radiation
fluid in the warm inflationary  scenario by means of an effective bulk
dissipative pressure and generalized the attractor condition. Likewise, we have
resorted to the synergistic mechanism to calculate the production of entropy
and the evolution of temperature. The exit temperature results lower but
approximately of the same order than the initial temperature. This renders
the reheating phase redundant.

Further, we have found that the combination of the synergistic mechanism and
the decay of the scalar fields into the matter fluid produces significant
entropy perturbations with proportional spectral amplitude and dominant mode
evolution to that of energy density perturbations on large wavelength scales
until inflation exit. This steep growth contrasts with other models of
inflation where long wavelength curvature modes evolve isentropically (see e.g.
\cite{gordon}, \cite{OJ}); however in our case there is a continuous transfer
of energy from the scalar fields to matter. Moreover, observational bounds on
the curvature perturbations at Hubble scale entry do not force on our model
any slow--roll constraints on the scalar field.

\section*{Acknowledgments}
LPC, ASJ and NZ thank the University of Buenos Aires for partial support under
project TX-93. This work was partially supported by the Spanish Ministry of
Science and Technology under grant BFM 2000--0351-C03--01.


\end{document}